\documentstyle[12pt,aasms4]{article}
\tighten

\newcommand\be{\begin{equation}}
\newcommand\ee{\end{equation}}

\input epsf
\lefthead{P. Kumar and S. Basu}
\righthead{Line asymmetry of solar p-modes}
\begin{document}
\baselineskip=15pt

\title{\bf Source depth for solar p-modes}

\author{Pawan Kumar\altaffilmark{1} and Sarbani Basu\altaffilmark{2}}

\altaffiltext{1}{Institute for Advanced Study, Olden Lane, Princeton, 
NJ 08540, U. S. A.}
\altaffiltext{2}{Astronomy Department, Yale University, P.O. Box 208101,
New Haven, CT 06520-8101, U. S. A.}

\begin{abstract}
Theoretically calculated power spectra are compares with
observed solar p-mode velocity power spectra 
over a range of mode degree and frequency. The depth for
the sources responsible for exciting p-modes of frequency 
$2.0$ mHz is determined from the
asymmetry of their power spectra and found to be about
800 km below the photosphere for quadrupole sources and 
150 km if sources are dipole. The source depth for 
high frequency oscillations of frequency greater than
$\sim$ 6 mHz is 180 (50) km for quadrupole (dipole) sources.

\end{abstract}


\section{Introduction}
During the last ten years there has been increasing evidence 
that solar p-modes are excited as a result of sound generated by 
turbulent convection (e.g., Goldreich et al. 1994; Georgobiani et al. 2000).
Acoustic oscillations with frequency above $\sim$5.5 mHz, the acoustic cutoff 
frequency in the solar atmosphere, provide good support for the general
validity of this mechanism (Kumar 1994). The 
observed asymmetry of p-mode line profiles 
and the opposite sense of asymmetry in the velocity and the intensity
power spectra have also been successfully modeled within the frame work
of the stochastic excitation theory (Abrams \& Kumar 1996; Roxburgh
\& Vorontsov 1997; Nigam et al. 1998; Kumar \& Basu 1999a).

Mode  excitation appears to be concentrated very close to the top
of the convection zone (cf. Kumar 1994, Kumar \& Basu 1999b), which is
consistent with the theory of convective excitation. According to
the mixing length theory of convection the location for the excitation
of higher frequency acoustic waves should be a little higher up in the
convection zone than lower frequency oscillations since the convective
frequency increases with radius. 

The theoretical modeling of line-asymmetry of low frequency p-modes
provides a means to determine the depth at which these oscillations are
excited. The source depth for high frequency oscillations ( $\nu  > 5.5$ mHz,
 the acoustic cutoff in the solar atmosphere) can be obtained 
from the frequency separation between adjacent peaks in the power spectrum.
The purpose of this paper is to determine the source depth for both high and 
low frequency acoustic oscillations and to compare it with the
depth expected from convective theory. This, we hope, will teach us
about convective velocity etc. in the top few scale heights of the
convection zone where the mixing length formalism is known to be very inaccurate.

\section{Theoretical calculation  of spectrum}

The calculation of power spectra is carried out using the
method described in Abrams \& Kumar (1996) and Kumar (1994).
Briefly, a coupled set of linearized mass,
momentum and entropy equations, with a source term,
is solved using the Green's function method. The source is parameterized
 by two numbers -- the depth where the source peaks and the radial
extent (the radial profile is taken to be a Gaussian function).
We calculate power spectra for dipole and quadrupole
sources using
\be
P(\omega) = \left| \int dr S(r,\omega)\; {d^n G_\omega \over dr^n}\right |^2,
\ee
where $n=0$ for dipole and $1$ for quadrupole sources, and $G_\omega$
is the Green's function for the linearized set of non-adiabatic wave equations.
Physically, dipole sources produce acoustic waves by applying a time
dependent force on the fluid. Only fluctuating internal stresses
are associated with quadrupole sources which have no associated
net momentum flux.

The power spectrum of p-modes below the acoustic cut-off frequency of 
$\sim 5.5$mHz is distinctly asymmetric and the magnitude of asymmetry 
is a function of source depth which has been used by Abrams \& Kumar (1996) 
and Kumar \& Basu (1999b) to determine the source depth for low frequency
modes. The frequency spacing between adjacent peaks above the
acoustic cutoff frequency can be used to determine the source depth
for high frequency waves (Kumar, 1994). These works provided preliminary
evidence for frequency dependence of source depth. In this paper
we carry out the fit to the best available observed  low and high frequency solar
spectra using an up to date. solar model, to refine 
the source depth determination as a function of wave frequency.

We use two solar models to calculate the theoretical power spectrum. 
One (``the standard model'') is a standard model of the present Sun.
It is constructed with OPAL opacities (Iglesias \& Rogers 1996) supplemented by
low temperature opacities of Kurucz(1991), and the OPAL
equation of state (Rogers, Swenson \& Iglesias 1996). Convective flux
is calculated using the formulation of Canuto \& Mazzitelli (1991), and
the photospheric structure is calculated using the empirical $T-\tau$
relation of Vernazza et al.~(1981). To check dependencies
on different ways of calculating convective transport
we have also used a similar model which uses the standard mixing length formalism
to calculate convective flux. The second model (``the old model'') is the
old Christensen-Dalsgaard model used in Kumar (1994) to determine
source depth for high frequency oscillation, and we use this model
to determine the error in source depth resulting from the use
of different solar models.

The observed power spectra used for the low frequency modes are the 360 
day data from the Michelson Doppler Imager (MDI) on board SoHO (Solar 
and Heliospheric Observatory); low frequency modes have small
line-width and require long time series to resolve their line profile.
For the high frequency waves we used data from GONG obtained during
months 10 to 22 of its operation. This period corresponds to a rough
minimum in solar activity. The monthly spectra were averaged together to
reduce noise.

\section{The fitting process}

\begin{figure}
\plotone{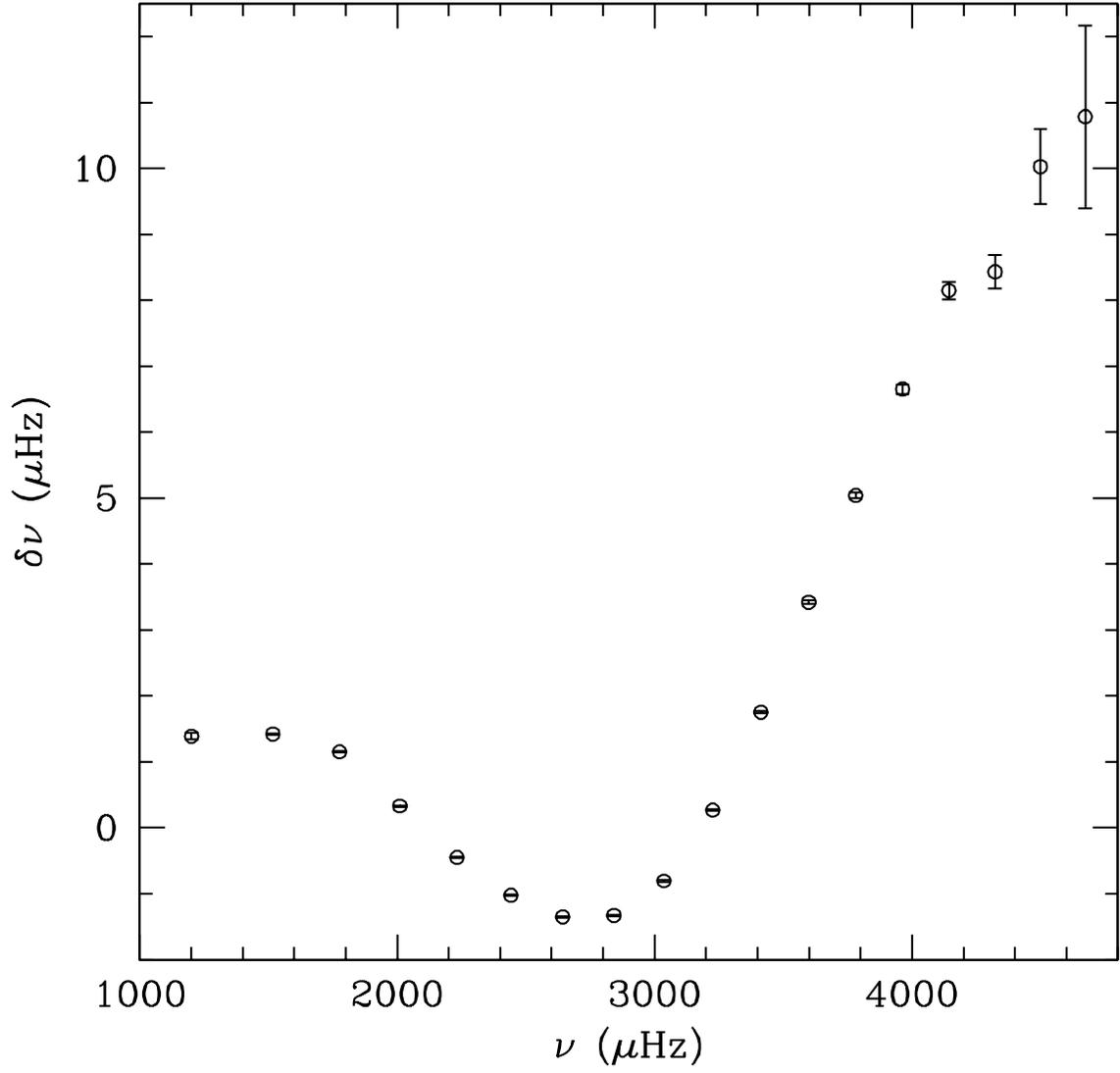}
\figcaption{The difference between the adiabatic frequencies of the standard 
solar model and the observed solar p-mode frequencies. Note that the
frequency difference increases rapidly at high frequencies.
\label{fig:freqdif}
}
\end{figure}

The fit to the high-frequency part of the power spectra contains four
parameters: the source depth, a constant background, an over all amplitude 
normalization factor (we normalize the amplitudes to 1 at 6.5 mHz)
and a uniform linear frequency shift of the power spectra which is needed
because the solar model we use is not perfect i.e., there is a frequency 
difference between the solar model and the Sun.
Figure~\ref{fig:freqdif} illustrates the frequency difference between the
standard model and the Sun. Since the
source depth determines the inter-peak frequency spacings rather than the
absolute frequency position of the peaks, this shift should not
affect the source depth determination.
The theoretically calculated power-spectra include the effect of 
$\ell$--leakage and Nyquist folding (the Nyquist frequency for
GONG data set is 8.333 mHz) before comparison with the observed
spectra is made. 

At low frequencies, we have the same number of free-parameters as 
Kumar \& Basu (1999b), i.e. the peak amplitude (which is normalized 
to unity), the line width, and the background (which is taken to be 
constant over the frequency range of the spectrum we model).
Unlike Kumar \& Basu (1999b), we take into account possible 
distortions of the power-spectra due to m-leakage. This is clearly seen
in very low frequency modes which have line-widths of the order of the
spacing between modes of $\delta m=\pm 1$.
The $\ell$-leakage is also important and is included in our theoretically
calculated spectra. The $\ell$-leakage is estimated from the ratio of different 
$\ell$-peaks at low frequencies of the observed spectrum; modes 
of degree $\ell\pm2$ contribute about 15\% of their power to the spectra
for modes of degree $\ell$ whereas the leakage from $\ell\pm1$ is $\sim50$\%.

The best-fit source depth is determined by minimizing the merit 
function (cf. Anderson, Duvall \& Jeffries 1990)
\be
F_m={1\over N} \sum_{i=1}^N\left({O_i-M_i}\over M_i\right)^2
\ee
where,  the summation is over all $N$  data points, $O_i$ is the observed power
and $M_i$ the theoretically calculated power. 

\section{Results}

A summary of the source-depths obtained with the standard solar model
for different cases is shown in Table~1.

\begin{figure}
\plotone{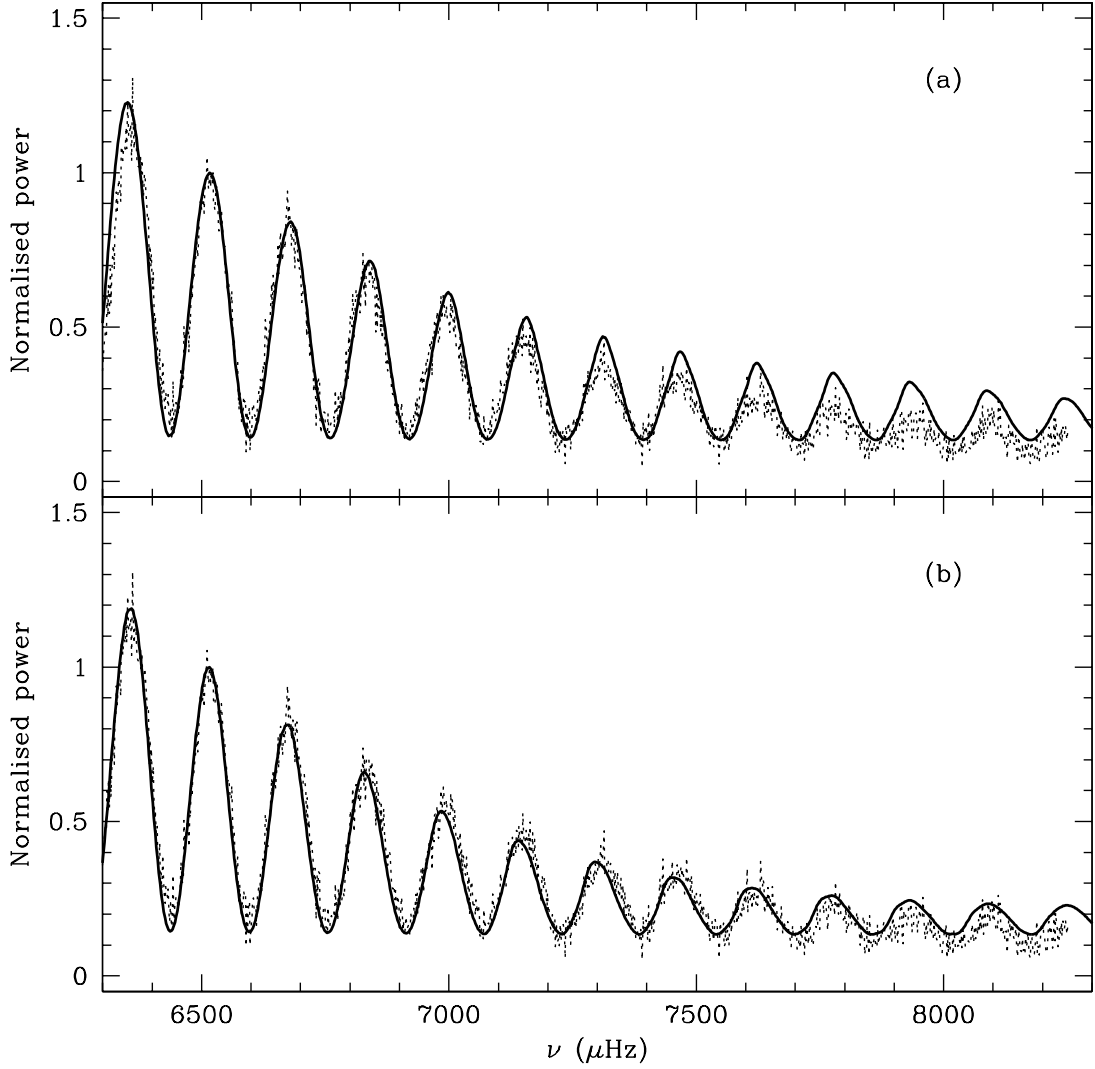}
\figcaption{Panel (a):  The best-fit quadrupole-source 
power-spectrum (continuous curve) superposed on the observed spectrum 
for $\ell=60$ (dotted curve). 
The source depth of the theoretical curve is 189 km and the frequency shift 
required is $-24$ $\mu$Hz. 
Panel (b):  The best-fit dipole-source power-spectrum (continuous curve)
superposed on the observed spectrum for $\ell=60$ (dotted curve). The
source depth of the theoretical curve
is 40 km and the theoretical spectrum was shifted by $-61$ $\mu$Hz in order to
fit the observed peaks. The radial extent of sources is $\sim 50$ km.
\label{fig:highquad}
}
\end{figure}

Figure~\ref{fig:highquad} shows the quadrupole and dipole fits to the
high frequency spectrum. The dipole sources seem
to provide as good a fit to the  observed spectrum, however they have
a serious drawback described below.

In order to fit the observed spectrum, the theoretical power 
spectrum for the dipole sources, placed at a depth of 30--130 km for
the best fit, has to be shifted by $-61\mu$Hz.
The spectrum calculated with quadrupole sources is shifted by a 
lower amount of $-24\mu$Hz in order to match the observed high
frequency peaks. The theoretically expected shift is about 
$-11$ $\mu$Hz at 4.3 mHz due to the inaccuracy of the solar model
we use. The difference between the theoretical and the observed 
p-mode frequencies increases rapidly with frequency (see fig. 1)
and a shift of $\sim20\mu$Hz at about 6 mHz is perhaps not unreasonable.
However, a shift of 61$\mu$Hz, needed for dipole excitation, is about
half the frequency spacing between peaks, and is much larger than 
can be accounted for as resulting from solar model error. No location for
dipole sources provides an acceptable fit to the high frequency
spectrum without a large frequency shift of the power spectrum and we 
therefore conclude that the wave excitation for high frequency waves
in the Sun is quadrupolar.

The quadrupole excitation requires a source depth of between 60--250 km
(the uncertainty being mainly a result of error in estimating the 
$\ell$-leakage and background removal); the depth is measured from 
the top of the convection zone. The old model gives similar source depths,
between 50 to 300 km for quadrupole source and 10 to 80 km for
dipole sources. The use of different convective theories to calculate the
mean solar model, such as the mixing length theory and the model
of Canuto \& Mazzitelli (1991), makes almost no difference to the
determination of source depth given the uncertainty in the
$\ell$-leakage.

The source-depth for the high-frequency waves is smaller than what 
Kumar \& Basu (1999b) found for low frequency p-modes of $\ell=35$ 
using the line asymmetry. We repeat their work for low frequency
p-modes of $\ell=$55 \& 60.

\begin{figure}
\plotone{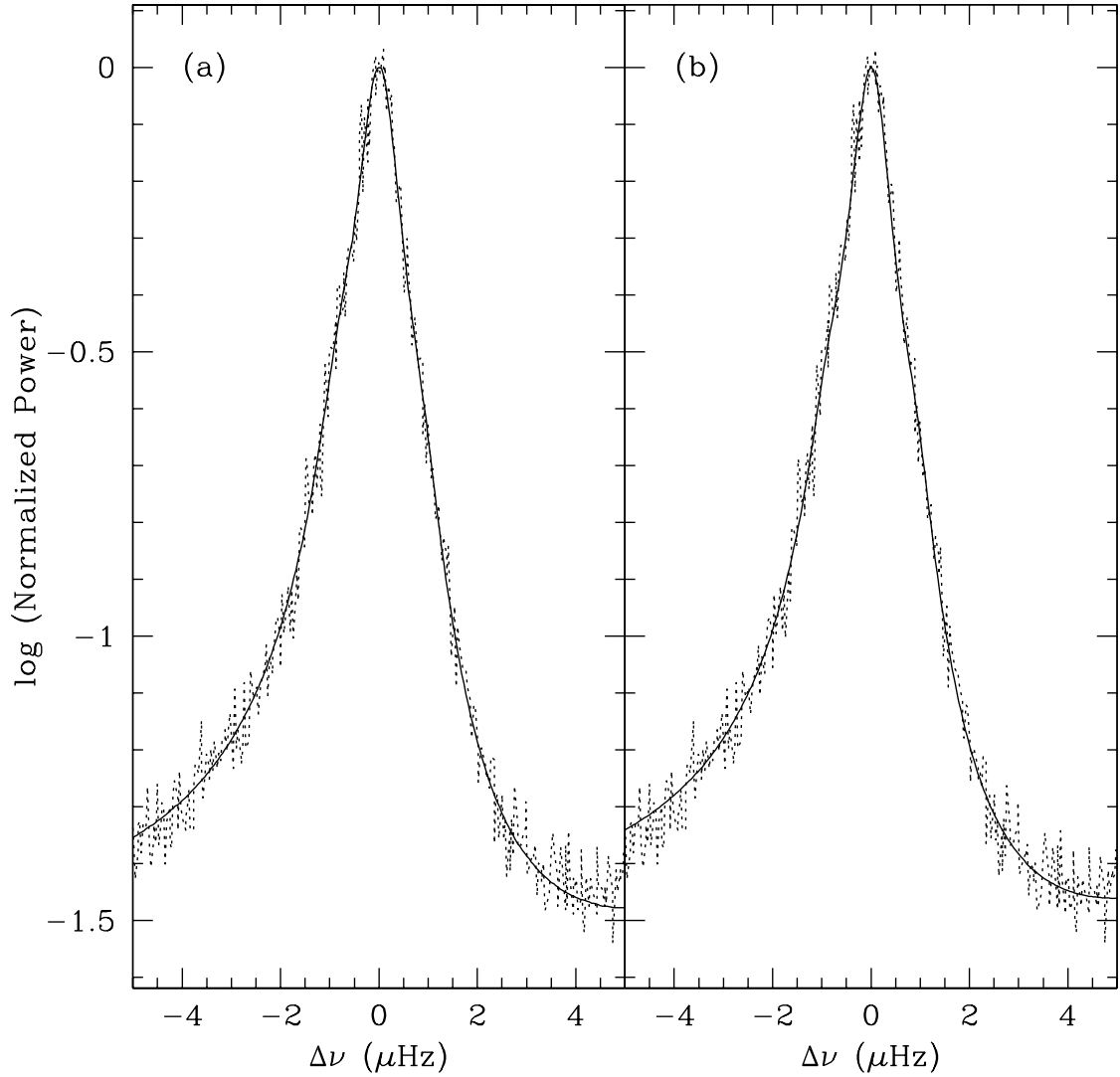}
\figcaption{Panel (a): Quadrupole-source power spectrum for 
an $\ell=60$ $\nu=2.01$mHz mode. The source depth of the theoretical
curve (continuous line) is 900 km. The dotted lines is the 
observed power spectrum. Panel (b):
The fit for dipole sources at a depth of 100 km.
The radial extent of sources is $\sim 100$ Km.
\label{fig:l60quad}
}
\end{figure}

Figure~\ref{fig:l60quad} shows the
theoretical power spectra for quadrupole and dipole sources
superposed on the observed spectrum of a mode with $\ell=60$,
radial order $n=4$, and frequency $\nu=2.01$mHz. The figure of merit per
degree of freedom for the best fit is 0.085 for quadrupolar as well as
dipole sources. Unlike the high frequency case, the source
depth required is quite different for the two source types.
The quadrupole sources have to be at depths of 700---1000 km
to match the observed line asymmetry, while the dipole sources have
to be placed at a depth of 120---350 km. These results are consistent
with those found by Kumar and Basu (1999) for
modes of $\ell=35$. The results for $\ell=55$ \& 60 are identical.
The standard model indicates that the  depth at which a 2 mHz mode starts 
to propagate is $\sim1500$ km.
Therefore, the excitation of the mode is taking place in the
evanescent region.

The main result of this work is that the source depth for low 
frequency p-modes (2 mHz) is much larger than the source location 
for acoustic waves of frequency larger than $\sim5.5$ mHz.

\bigskip
\section{Discussion}
\medskip

The depth dependence of energy input rate to a p-mode due to quadrupole
excitation is given by (Goldreich et al. 1994)

$$
{d\dot E\over dr}\sim {2\pi\omega_\alpha^2\over 5}
r^2\rho^2v_{\Lambda}^3 \left|{\partial\xi_\alpha^r \over\partial
r}\right|^2 {\Lambda^4 \left({\cal R}^2 +1\right){\cal S}^2
\over 1+ (\omega_\alpha\tau_{\Lambda}/\eta)^{15/2}}, \eqno(3)
$$
where $\omega_\alpha$ is mode frequency, $\Lambda$ is vertical coherence
length of largest eddies which in the mixing length theory is
a constant multiple of pressure scale height, $v_{\Lambda}$ is the
convective velocity, $\tau_\Lambda=\Lambda/v_\Lambda$ is the eddy turnover 
time, $\xi^r_\alpha$ is the normalized
radial displacement eigenfunction for the mode, $\eta$ is a constant
factor of order $\pi/2$, ${\cal R}$ is the dimensionless compressibility
of the mode which is close to unity for high order p-modes, and ${\cal S}$
is the horizontal coherence length of eddies and is taken to be same for
all inertial range eddies.

Figure 4 shows this function for modes of frequency 1.7, 2.1 and 5 mHz 
for a standard solar model and the mixing length theory of convection.
Note that the depth for 1.7 mHz mode is about 400 km below the
photosphere whereas the peak excitation depth for 5 mHz waves
is $\sim50$ km. Moreover, these depths are much smaller than what 
we obtained in the last section for the preferred quadrupole sources 
by the modeling of line asymmetry for
low frequency modes and the frequency spacing between adjacent peaks 
in power spectra at high frequencies. This indicates some error in the
mixing length model of convective in the top few scale heights.

A possible ad hock modification of the convective profile to yield
larger source depth is to decrease the convective timescale $\tau_\Lambda$
in the top few scale heights of the convection zone.
Since $\tau_\Lambda$ enters as a very high power in equation (3) most
of the excitation is concentrated at the resonance layer, i.e. where
$\tau_\Lambda\sim \omega_\alpha^{-1}$, and a modest decrease in $\tau_\Lambda$
is sufficient to reconcile the ``observed'' and the theoretical source
depths. For instance to shift the quadrupolar excitation depth for 1.7mHz mode 
to 1000 km below the photosphere we need to decrease $\tau_\Lambda$ by a
factor of about 2 in the top 1000 km of the convection zone; this shifts 
the peak excitation depth for 2.0 mHz mode to 900 km and the high 
frequency acoustic waves to $\sim 150$ km. A decrease in $\tau_\Lambda$
by a factor of 1.5 causes most of the excitation for 2.0 mHz mode
to occur at a depth of $\sim$ 700 km. A decrease in $\tau_\Lambda$ could
be caused by an increase in the convective velocity or a decrease in the
radial coherence length for turbulent eddies or a combination of the two.
The numerical simulation of solar convection (Abbett et al. 1997)
indicates that the convective velocity in the top 2000 km
is larger than expected from MLT by a factor of $\sim 1.5$.
We should not expect a modification to $\tau_\Lambda$ to be a constant
factor independent of $r$. In fact the "observed" depth for low
and high frequency waves does not seem to be compatible with a 
constant, depth independent, decrease to $\tau_\Lambda$, but instead
suggests that the modification factor is larger at smaller depths.

We expect realistic numerical simulations of the solar 
convection zone, such as those  carried out by groups in Copenhagen, Michigan
and Yale,
to provide a deeper understanding for the deficiencies in the convective 
modeling of the superadiabatic layer near the top of the solar convection zone
and shed light on the underlying physical cause for the greater source 
depth that we find.

\acknowledgments

This work utilizes data obtained by the Global Oscillation Network
Group (GONG) project, managed by the National Solar Observatory.
This work also utilizes data from the Solar Oscillations
Investigation / Michelson Doppler Imager (SOI/MDI) on the Solar
and Heliospheric Observatory (SOHO).  SOHO is a project of
international cooperation between ESA and NASA.
This work is supported by a NASA grant NAG 5-8328.

\clearpage
\begin{table}
\caption{Source depths}
\begin{tabular}{ccc}
\hline
\noalign{\smallskip}
Frequency & \multicolumn{2}{c}{Source depth (km)} \cr
Range & Quadrupole & Dipole \cr
\noalign{\smallskip}
\hline
\noalign{\smallskip}
$\ell=60$, $\nu > 6000\;m$Hz&  60 -- 250 & 30 -- 130 \cr
$\ell=60$, $\nu=2.01\;m$Hz & 700 -- 1000 & 120 -- 350 \cr
$\ell=55$, $\nu=1.72\;m$Hz & 700 -- 1000 & 120 -- 350 \cr
\noalign{\smallskip}
\hline
\end{tabular}
\end{table}

\clearpage

\begin{figure}
\plotone{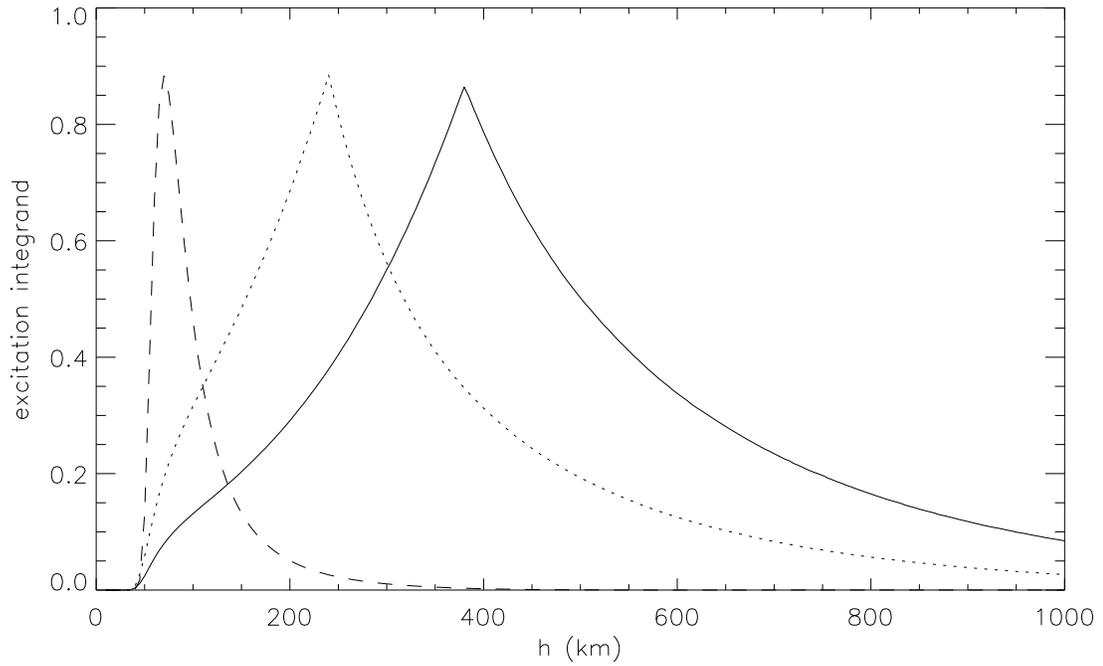}
\figcaption{The rate of energy input in p-modes as a function of depth
for quadrupolar sources (expression in eq. 3) is shown for three p-modes
of frequency 1.7 mHz (continuous line), 2.1 (dotted curve) and 5.0 mHz
(dashed line) all of degree $\ell=55$. The convective velocity in
the solar model was calculated using the standard mixing length theory.
\label{fig:fig4}
}
\end{figure}

\end{document}